# An extended isogeometric analysis for vibration of cracked FGM plates using higher-order shear deformation theory

Loc V. Tran[1], Vinh Phu Nguyen[2], M. Abdel Wahab[3], H. Nguyen-Xuan[1]

[1]Division of Computational Mechanics, Ton Duc Thang University Ho Chi Minh City, Vietnam

[2]Cardiff University, Queen's Buildings, The Parade, Cardiff CF24 3AA, Wales, UK

[3]Department of Mechanical Construction and Production, Faculty of Engineering and Architecture, Ghent University, 9000, Ghent – Belgium

**Abstract**

A novel and effective formulation that combines the eXtended IsoGeometric Approach (XIGA) and Higher-order Shear Deformation Theory (HSDT) is proposed to study the free vibration of cracked Functionally Graded Material (FGM) plates. Herein, the general HSDT model with five unknown variables per node is applied for calculating the stiffness matrix without needing Shear Correction Factor (SCF). In order to model the discontinuous and singular phenomena in the cracked plates, IsoGeometric Analysis (IGA) utilizing the Non-Uniform Rational B-Spline (NURBS) functions is incorporated with enrichment functions through the partition of unity method. NURBS basis functions with their inherent arbitrary high order smoothness permit the $C^1$ requirement of the HSDT model. The material properties of the FGM plates vary continuously through the plate thickness according to an exponent function. The effects of gradient index, crack length, crack location, length to thickness on the natural frequencies and mode shapes of simply supported and clamped FGM plate are studied. Numerical examples are provided to show excellent performance of the proposed method compared with other published solutions in the literature.

# 1. Introduction

Based on the characteristics of the bamboo, in 1984, a group of scientists in Sendai-Japan proposed a new smart material, i.e. the so–called Functionally Graded Material (FGM) [1-4] with a continuous change of material property along certain dimensions of the structure. FGM is often a mixture of two distinct material phases: e.g. ceramic and metal with the variation of the volume fraction according to power law through the thickness. As a result, FGMs are enabled to inherit the best properties of the components, e.g. low thermal conductivity, high thermal resistance by ceramic and ductility, durability of metal. They are therefore more suitable to use in aerospace structure applications and nuclear plants, etc.

In order to use them efficiently, a clear understanding of their behaviors such as deformable characteristic, stress distribution, natural frequency and critical buckling load under various conditions is required. Hence, investigation on property of FGM structure has been addressed since long time. For instance, Reddy [5] proposed an analytical formulation based on a Navier's approach using the third-order shear deformation theory and the von Kármán-type geometric non-linearity. Vel and Battra [6,7] introduced an exact formulation based on the form of a power series for thermo-elastic deformations and vibration of rectangular FGM plates. Yang and Shen [8] have analyzed the dynamic response of thin FGM plates subjected to impulsive loads. Cinefra et al. [9] investigated the response of FGM shell structure under mechanical load. Nguyen et al. [10-12] studied the behaviors of FGM using smoothed finite element method. Ferreira et al. [13,14] performed static and dynamic analysis of FGM plate based on higher-order shear and normal deformable plate theory using the meshless local Petrov–Galerkin method. Tran et al. [15] studied the thermal buckling of FGM plate based on third-order shear deformation theory.

From the literature, these works are carried out for designing the FGM plate structures without the presences of cracks or flaws. However, during manufacturing the FGM or general plate structures may have some flaws. In service, from the flaws, the cracks are generated and grown when subjected to large cyclic loading. They lead to a reduction of the load carrying capacity of the structures. To clearly understand the dynamic behavior of FGM plate with initial cracks is thus too necessary. Vibration of cracked plates was early studied in 1967 using Green's function approach by Lynn and Kumbasar [16]. Later, Stahl and Keer [17] studied the vibration of cracked rectangular plates using Levy-Nadai approach. Instead of taking into account the special treatment to consider the singular phenomenon at the crack tips like Stahl and Keer, Liew et al. [18] used domain decomposition method to devise the plate domain into the numerous subdomains around the crack location. Other numerical methods used to study cracked plates are: finite Fourier-series transform [19], Rayleigh–Ritz method [20], harmonic balance method [21], and Finite Element Method (FEM) [22, 23], eXtended Finite Element Method (XFEM) [24]. Among them, XFEM originated by Belytschko and Black [25] is known as a robust numerical method, which uses enrichment functions to model discontinuities independent of the finite element mesh. Recently, Natarajan et al. [26] extended this method for dynamic analysis of FGM plate based on the FSDT model. This model is simple to implement and is applicable to both thick and thin FGM plates. However, the accuracy of solutions will be strongly dependent on the Shear Correction Factors (SCF) of which their values are quite dispersed through each problem, e.g. SCF is equal to 5/6 in Ref.[27], $\pi^2/12$ in Ref.[28] or a complicated function derived from equilibrium condition [29].

In this paper, we develop the HSDT model that includes higher-order terms in the approximation of the displacement field for modeling FGM plate. It is worth mentioning that this model requires $C^1$-continuity of the generalized displacements leading to the second-order derivative of the stiffness formulation which causes some obstacles in standard $C^0$ finite formulations. Fortunately, it is shown that such a $C^1$-HSDT formulation can be easily achieved using a NURBS-based isogeometric approach [30, 31]. In addition, to capture the discontinuous phenomenon in the cracked FGM plates, the enrichment functions through Partition of Unity Method (PUM) is incorporated with NURBS basic functions to create a novel method as so-called eXtended Isogeometric Analysis (XIGA) [32]. Herein, our study focuses on investigating the vibration of the cracked FGM plate with an initial crack emanating from an edge or centrally located.

Several numerical examples are given to show the performance of the proposed method and results obtained are compared to other published methods in the literature.

The paper is outlined as follows. The next section introduces the governing equation for FGM plate based on HSDT model. In section 3, an incorporated method between the enrichment functions through PUM and IGA – based NURBS function are used to simulate the cracked FGM plates. Numerical results and discussions are provided in section 4. Finally, the article is closed with some concluding remarks.

## 2. Governing equations for functionally graded plates

*2.1. Functionally graded material*

Functionally graded material is a composite material which is created by mixing two distinct material phases. Two mixed materials are often ceramic at the top and metal at the bottom as shown in Figure 1. In our work, two homogenous models have been used to estimate the effective properties of the FGM include the rule of mixture [5] and the Mori-Tanaka technique [33]. Herein, the volume fraction of the ceramic and metal phase is described by the following power-law exponent function

$$V_c(z) = \left(\frac{1}{2} + \frac{z}{h}\right)^n, \quad V_m = 1 - V_c \tag{1}$$

where subscripts *m* and *c* refer to the metal and ceramic constituents, respectively. Eq. (1) implies that the volume fraction varies through the thickness based on the power index *n*.

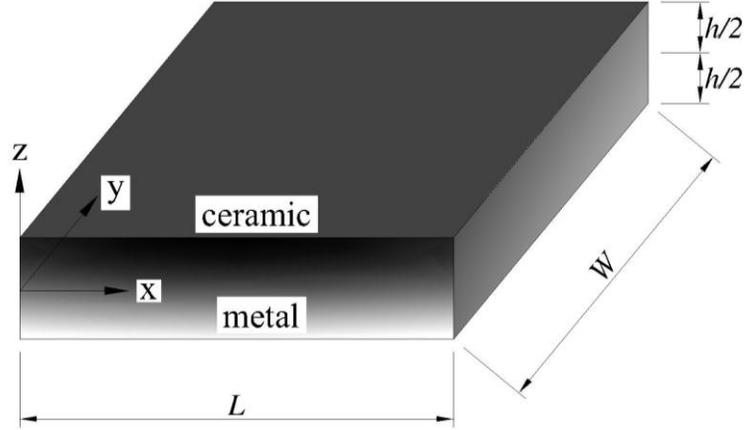

Figure 1: The functionally graded plate model.

The effective material properties according to the rule of mixture are the given by

$$P_e = P_c V_c + P_m V_m \qquad (2)$$

where $P_c$ and $P_m$ denote the material properties of the ceramic and the metal, respectively, including the Young's modulus $E$, Poisson's ratio $\nu$ and the density $\rho$.

However, the rule of mixture does not consider the interactions among the constituents [34]. So, the Mori-Tanaka technique [33] is then used to take into account these interactions with the effective bulk and shear modulus defined from following relations:

$$\frac{K_e - K_m}{K_c - K_m} = \frac{V_c}{1 + V_m \frac{K_c - K_m}{K_m + 4/3\mu_m}}$$
$$\frac{\mu_e - \mu_m}{\mu_c - \mu_m} = \frac{V_c}{1 + V_m \frac{\mu_c - \mu_m}{\mu_m + f_1}} \qquad (3)$$

where $f_1 = \dfrac{\mu_m(9K_m + 8\mu_m)}{6(K_m + 2\mu_m)}$. And the effective values of Young's modulus $E$ and Poisson's ratio $\nu$ are given by

$$E_e = \frac{9K_e \mu_e}{3K_e + \mu_e}, \qquad \nu_e = \frac{3K_e - 2\mu_e}{2(3K_e + \mu_e)} \qquad (4)$$

Figure 2 illustrates comparison of the effective Young's modulus of Al/ZrO$_2$ FGM plate calculated by the rule of mixture and the Mori-Tanaka scheme via the power index *n*. Note that with homogeneous material, the two models produce the same values. For inhomogeneous material, the effective property through the thickness of the former is higher than that of latter. Moreover, increasing in power index *n* leads to decrement of the material property due to the rise of metallic volume fraction.

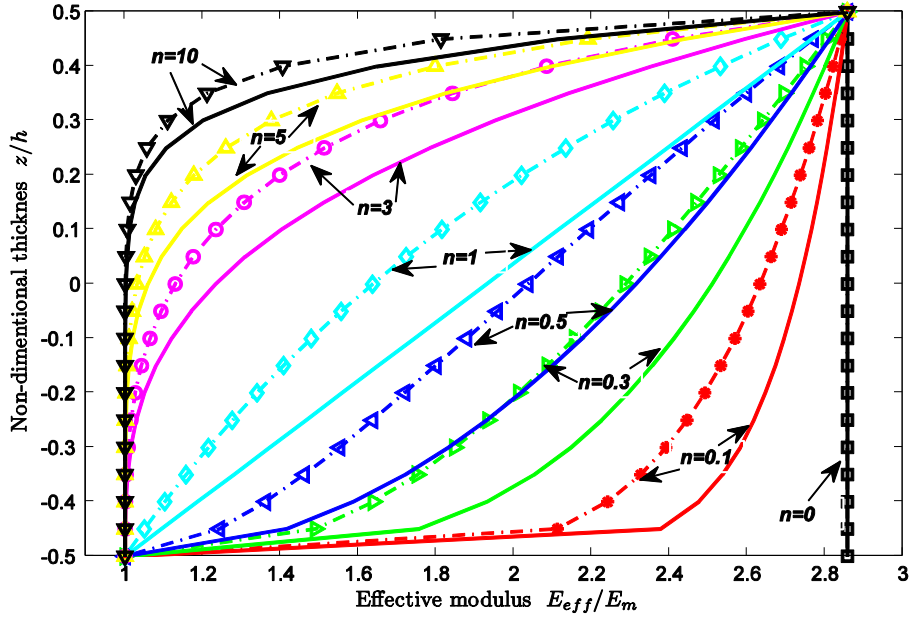

Figure 2. The effective modulus of FGM plate computed by the rule of mixture (in solid line) and the Mori-Tanaka (in dash dot line).

*2.2. General plate theory*

To consider the effect of shear deformation directly, the generalized five-parameter displacement field based on higher-order shear deformation theory is defined as

$$\mathbf{u} = \mathbf{u}_1 + z\mathbf{u}_2 + f(z)\mathbf{u}_3 \qquad (5)$$

where $\mathbf{u}_1 = \{u_0 \ v_0 \ w\}^T$ is the axial displacement, $\mathbf{u}_2 = -\{w_{,x} \ w_{,y} \ 0\}^T$ and $\mathbf{u}_3 = \{\beta_x \ \beta_y \ 0\}^T$ are the rotations in the *x*, *y* and *z* axes, respectively. $f(z)$ is the so-called distributed function which is chosen to satisfy the tangential zero value at the plate surfaces, i.e. $f'(\pm h/2) = 0$. Based on this condition, various distributed functions $f(z)$ have been proposed: third-order polynomials by Reddy [35], exponential function by Karama [36], sinusoidal function by Arya [37], fifth-order polynomial by Nguyen [38] and inverse tangent functions by Thai [39] as shown in Figure 3.

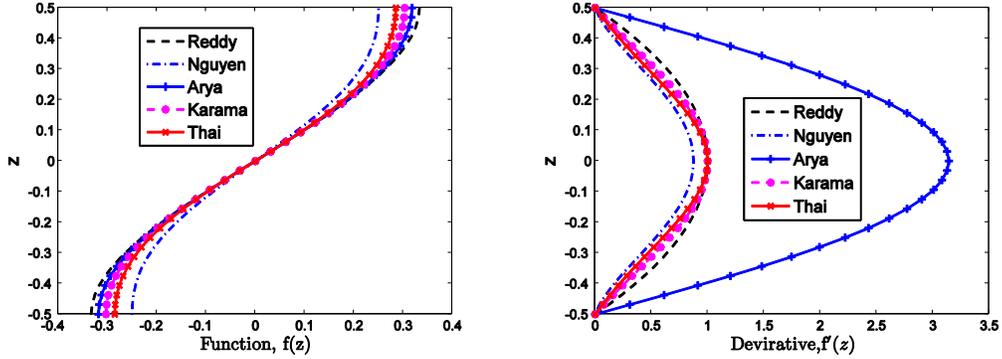

Figure 3. The shape functions and their derivative through plate thickness.

Here, we chose the third-order shear deformation theory (TSDT) [35] because of popularity by setting $f(z) = z - \frac{4}{3h^2}z^3$. Moreover, by setting $f(z) = z$ and substituting $\phi_\mathbf{x} = -w_{,\mathbf{x}} + \beta_\mathbf{x}$ in Eq.(5), the first order shear deformation theory (FSDT) is obtained

$$\begin{aligned} u(x,y,z) &= u_0 + z\phi_x \\ v(x,y,z) &= v_0 + z\phi_y \\ w(x,y,z) &= w \end{aligned} \quad (6)$$

The strains of the mid-surface deformation are given by

$$\begin{Bmatrix} \boldsymbol{\varepsilon} \\ \boldsymbol{\gamma} \end{Bmatrix} = \begin{Bmatrix} \boldsymbol{\varepsilon}_0 + z\boldsymbol{\kappa}_1 + f(z)\boldsymbol{\kappa}_2 \\ f'(z)\boldsymbol{\beta} \end{Bmatrix} \quad (7)$$

where

$$\boldsymbol{\varepsilon}_0 = \begin{bmatrix} u_{0,x} \\ v_{0,y} \\ u_{0,y} + v_{0,x} \end{bmatrix}, \quad \boldsymbol{\kappa}_1 = -\begin{bmatrix} w_{,xx} \\ w_{,yy} \\ 2w_{,xy} \end{bmatrix}, \quad \boldsymbol{\kappa}_2 = \begin{bmatrix} \beta_{x,x} \\ \beta_{y,y} \\ \beta_{x,y} + \beta_{y,x} \end{bmatrix}, \quad \boldsymbol{\beta} = \begin{bmatrix} \beta_x \\ \beta_y \end{bmatrix} \quad (8)$$

From Eq.(7), it can be seen that the shear stresses vanish at the top and bottom surfaces of plate.

Using the Hamilton principle, the weak form for free vibration analysis of a FGM plate can be expressed as:

$$\int_\Omega \delta\boldsymbol{\varepsilon}^T \mathbf{D}^b \boldsymbol{\varepsilon} \, d\Omega + \int_\Omega \delta\boldsymbol{\gamma}^T \mathbf{D}^s \boldsymbol{\gamma} \, d\Omega = \int_\Omega \delta\tilde{\mathbf{u}}^T \mathbf{m} \ddot{\tilde{\mathbf{u}}} \, d\Omega \tag{9}$$

where

$$\mathbf{D}^b = \begin{bmatrix} \mathbf{A} & \mathbf{B} & \mathbf{E} \\ \mathbf{B} & \mathbf{D} & \mathbf{F} \\ \mathbf{E} & \mathbf{F} & \mathbf{H} \end{bmatrix} \tag{10}$$

in which

$$A_{ij}, B_{ij}, D_{ij}, E_{ij}, F_{ij}, H_{ij} = \int_{-h/2}^{h/2} (1, z, z^2, f(z), zf(z), f^2(z)) Q_{ij} \, dz$$

$$D_{ij}^s = \int_{-h/2}^{h/2} [f'(z)]^2 G_{ij} \, dz \tag{11}$$

the material matrices are given as

$$\mathbf{Q} = \frac{E_e}{1-\nu_e^2} \begin{bmatrix} 1 & \nu_e & 0 \\ \nu_e & 1 & 0 \\ 0 & 0 & (1-\nu_e)/2 \end{bmatrix}, \quad \mathbf{G} = \frac{E_e}{2(1+\nu_e)} \begin{bmatrix} 1 & 0 \\ 0 & 1 \end{bmatrix} \tag{12}$$

Herein, the mass matrix **m** is calculated according to consistent form as follow

$$\mathbf{m} = \begin{bmatrix} I_1 & I_2 & I_4 \\ I_2 & I_3 & I_5 \\ I_4 & I_5 & I_6 \end{bmatrix} \text{ with } I_i = \int_{-h/2}^{h/2} \rho_e \left(1, z, z^2, f(z), zf(z), (f(z))^2\right) dz \tag{13}$$

and

$$\tilde{\mathbf{u}} = \{\mathbf{u}_1 \ \mathbf{u}_2 \ \mathbf{u}_3\}^T \tag{14}$$

## 3. An extended isogeometric cracked plate formulation

*3.1. B-Spline/NURBS basic functions*

A knot vector $\Xi = \{\xi_1, \xi_2, ..., \xi_{n+p+1}\}$ is defined as a sequence of knot value $\xi_i \in R$, $i = 1,...n+p$. If the first and the last knots are repeated $p+1$ times, the knot vector is called an open knot. A B-spline basis function is $C^\infty$ continuous inside a knot span and $C^{p-1}$ continuous at a single knot. Thus, as $p \geq 2$ the present approach always satisfies $C^1$ requirement in approximate formulations of HSDT.

The B-spline basis functions $N_{i,p}(\xi)$ are defined by the following recursion formula

$$N_{i,p}(\xi) = \frac{\xi - \xi_i}{\xi_{i+p} - \xi_i} N_{i,p-1}(\xi) + \frac{\xi_{i+p+1} - \xi}{\xi_{i+p+1} - \xi_{i+1}} N_{i+1,p-1}(\xi)$$

$$\text{as } p = 0, \ N_{i,0}(\xi) = \begin{cases} 1 & if \ \xi_i < \xi < \xi_{i+1} \\ 0 & \text{otherwise} \end{cases}$$

(15)

By the tensor product of basis functions in two parametric dimensions $\xi$ and $\eta$ with two knot vectors $\Xi = \{\xi_1, \xi_2, ..., \xi_{n+p+1}\}$ and $\mathbf{H} = \{\eta_1, \eta_2, ..., \eta_{m+q+1}\}$, the two-dimensional B-spline basis functions are obtained

$$N_A(\xi, \eta) = N_{i,p}(\xi) M_{j,q}(\eta) \tag{16}$$

Figure 4 illustrates the set of one-dimensional and two-dimensional B-spline basis functions according to open uniform knot vector $\Xi = \{0, 0, 0, 0, 0.5, 1, 1, 1, 1\}$.

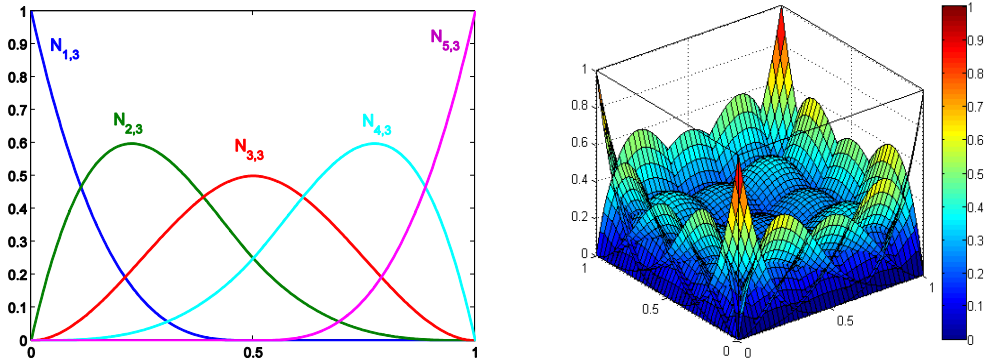

Figure 4. 1D and 2D B-spline basis functions.

To present exactly some curved geometries (e.g. circles, cylinders, spheres, etc.) the Non-Uniform Rational B-Splines (NURBS) functions are used. Be different from B-spline, each control point of NURBS has additional value called an individual weight $w_A$ [30]. Then the NURBS functions can be expressed as

$$R_A(\xi,\eta) = \frac{N_A w_A}{\sum_A^{m \times n} N_A(\xi,\eta) w_A} \quad (17)$$

It can be noted that B-spline function is only a special case of the NURBS function when the individual weight of control point is constant.

*3.2. Extended isogeometric finite element method*

In 1999, Ted Belytschko et al. [25] proposed the extended finite element method (XFEM) to analyze the linear elastic fracture mechanics problems without based on the mesh. The key idea is adding the new enriched functions to capture the local discontinuous and singular fields as follow:

$$\mathbf{u}^h(\mathbf{x}) = \sum_{I \in S} N_I(\mathbf{x}) \mathbf{q}_I^{std} + \textit{enrichment fields} \quad (18)$$

where $N_I(\mathbf{x})$ and $\mathbf{q}_I^{std} = \begin{bmatrix} u_{0I} & v_{0I} & w_{0I} & \beta_{xI} & \beta_{yI} \end{bmatrix}^T$ are the standard finite element shape function and nodal degrees of freedom associated with node *I*. To enhance the capability of IGA in analyzing cracked structures, a new numerical procedure – so-called eXtended

IsoGeometric Analysis (XIGA) is proposed by Luycker et al. [40], Ghorashi et al. [32], Nguyen et al. [41] as combination of IGA and PUM. Being different from XFEM, XIGA utilizes the NURBS basis functions instead of the Lagrange polynomials in approximation of the displacement field

$$\mathbf{u}^h(\mathbf{x}) = \sum_{I \in S} R_I(\xi) \mathbf{q}_I^{std} + \sum_{J \in S^{enr}} R_J^{enr}(\xi) \mathbf{q}_J^{enr} \tag{19}$$

in which $R_J^{enr}$ are the enrichment functions associated with node $J$ located in enriched domain $S^{enr}$ which is splitted up two parts including: a set $S^c$ for Heaviside enriched control points and a set $S^f$ for crack tip enriched control points as shown in Figure 5. Furthermore, the enriched functions determined specifically for each domain will be shown next.

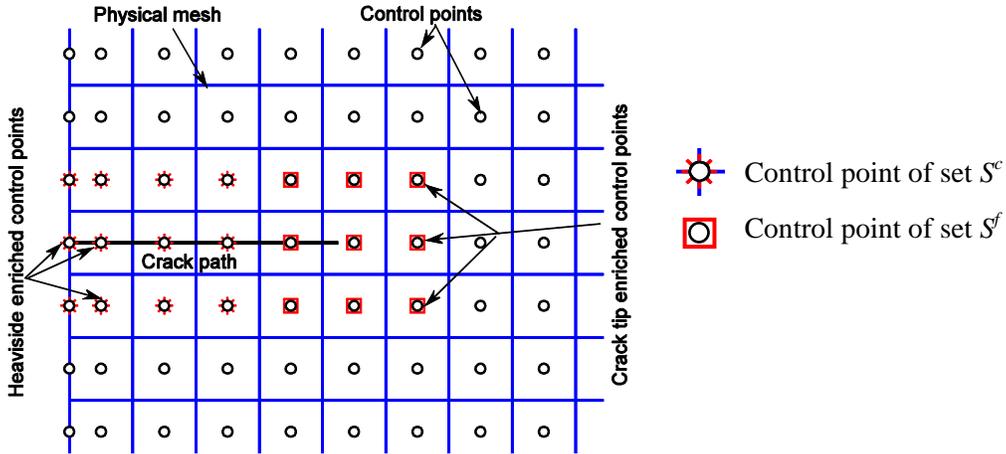

Figure 5. Illustration of the nodal sets $S^c$, $S^f$ for a quadratic NURBS mesh.

### 3.3. Cracked plate formulation based on HSDT

To describe the discontinuous displacement field, the enrichment function is given by

$$R_J^{enr}(\xi) = R_J(\xi)\big(H(\mathbf{x}) - H(\mathbf{x}_J)\big), \quad J \in S^c \tag{20}$$

where Heaviside function is defined as

$$H(\mathbf{x}) = \begin{cases} +1 & \text{if } (\mathbf{x}-\mathbf{x}^*)\mathbf{n} > 0 \\ -1 & \text{orthewise} \end{cases} \tag{21}$$

in which $\mathbf{x}^*$ is projection of point $\mathbf{x}$ on the crack path, and $\mathbf{n}$ is the normal vector of crack at point $\mathbf{x}^*$ (see more details in Ref [24]). And the singularity field near crack tip is modified by the branching functions [24] as follow

$$R_J^{enr}(\xi) = R_J(\xi)\left(\sum_{L=1}^{4}\left(G_L(r,\theta) - G_L(r_J,\theta_J)\right)\right), \quad J \in S^f \tag{22}$$

where

$$G_L(r,\theta) = \begin{cases} r^{3/2}\left[\sin\dfrac{\theta}{2} \quad \cos\dfrac{\theta}{2} \quad \sin\dfrac{3\theta}{2} \quad \cos\dfrac{3\theta}{2}\right] & \text{for } \mathbf{u}_1 \text{ variable} \\ r^{1/2}\left[\sin\dfrac{\theta}{2} \quad \cos\dfrac{\theta}{2} \quad \sin\dfrac{\theta}{2}\sin\theta \quad \cos\dfrac{\theta}{2}\cos\theta\right] & \text{for } \boldsymbol{\beta} \text{ variable} \end{cases} \tag{23}$$

in which $r$ and $\theta$ are polar coordinates in the local crack tip coordinate system.

Now, by substituting the displacement field approximated in Eq. (19) into Eq. (8) the strain matrices including in-plane and shear strains can be rewritten as:

$$\left[\boldsymbol{\varepsilon}_0^T \; \boldsymbol{\kappa}_1^T \; \boldsymbol{\kappa}_2^T \; \boldsymbol{\varepsilon}_s^T\right]^T = \sum_{I=1}^{m\times n}\left[\left(\mathbf{B}_I^m\right)^T \; \left(\mathbf{B}_I^{b1}\right)^T \; \left(\mathbf{B}_I^{b2}\right)^T \; \left(\mathbf{B}_I^s\right)^T\right]^T \mathbf{q}_I \tag{24}$$

in which the unknown vector $\mathbf{q}$ contains both displacements and enriched DOFs, and

$$\mathbf{B} = \left[\mathbf{B}^{std} \,\middle|\, \mathbf{B}^{enr}\right] \tag{25}$$

where $\mathbf{B}^{std}$ and $\mathbf{B}^{enr}$ are the standard and enriched strain matrices of $\mathbf{B}$ defined in the following forms:

$$\mathbf{B}^m = \begin{bmatrix} \bar{R}_{,x} & 0 & 0 & 0 & 0 \\ 0 & \bar{R}_{,y} & 0 & 0 & 0 \\ \bar{R}_{,y} & \bar{R}_{,x} & 0 & 0 & 0 \end{bmatrix}, \mathbf{B}^{b1} = -\begin{bmatrix} 0 & 0 & \bar{R}_{,xx} & 0 & 0 \\ 0 & 0 & \bar{R}_{,yy} & 0 & 0 \\ 0 & 0 & 2\bar{R}_{,xy} & 0 & 0 \end{bmatrix}, \tag{26}$$

$$\mathbf{B}^{b2} = \begin{bmatrix} 0 & 0 & 0 & \bar{R}_{,x} & 0 \\ 0 & 0 & 0 & 0 & \bar{R}_{,y} \\ 0 & 0 & 0 & \bar{R}_{,y} & \bar{R}_{,x} \end{bmatrix}, \quad \mathbf{B}^{s} = \begin{bmatrix} 0 & 0 & \bar{R}_{,x} & 0 & 0 \\ 0 & 0 & \bar{R}_{,y} & 0 & 0 \end{bmatrix}$$

where $\bar{R}$ can be either the NURBS basic functions $R(\xi)$ or enriched functions $R^{enr}$.

Substituting Eq. (7) with relation in Eq. (24) into Eq. (9), the formulations of free vibration problem can be rewritten as follow:

$$\left(\mathbf{K} - \omega^2 \mathbf{M}\right)\mathbf{d} = \mathbf{0} \tag{27}$$

where $\omega \in \mathbb{R}^+$ are the natural frequency and the global stiffness matrix $\mathbf{K}$ is given by:

$$\mathbf{K} = \int_\Omega \{\mathbf{B}^m \ \mathbf{B}^{b1} \ \mathbf{B}^{b2}\} \mathbf{D}^b \{\mathbf{B}^m \ \mathbf{B}^{b1} \ \mathbf{B}^{b2}\}^T + \mathbf{B}^{sT} \mathbf{D}^s \mathbf{B}^s \mathrm{d}\Omega \tag{28}$$

The global mass matrix $\mathbf{M}$ is expressed as:

$$\mathbf{M} = \int_\Omega \tilde{\mathbf{N}}^T \mathbf{m} \tilde{\mathbf{N}} \mathrm{d}\Omega \tag{29}$$

in which

$$\tilde{\mathbf{N}} = \begin{Bmatrix} \mathbf{N}_1 \\ \mathbf{N}_2 \\ \mathbf{N}_3 \end{Bmatrix}, \quad \mathbf{N}_1 = \begin{bmatrix} \bar{R} & 0 & 0 & 0 & 0 \\ 0 & \bar{R} & 0 & 0 & 0 \\ 0 & 0 & \bar{R} & 0 & 0 \end{bmatrix};$$

$$\mathbf{N}_2 = -\begin{bmatrix} 0 & 0 & \bar{R}_{,x} & 0 & 0 \\ 0 & 0 & \bar{R}_{,y} & 0 & 0 \\ 0 & 0 & 0 & 0 & 0 \end{bmatrix}; \quad \mathbf{N}_3 = \begin{bmatrix} 0 & 0 & 0 & \bar{R} & 0 \\ 0 & 0 & 0 & 0 & \bar{R} \\ 0 & 0 & 0 & 0 & 0 \end{bmatrix} \tag{30}$$

It is observed from Eq. (28) that the shear correction factors are no longer required in the stiffness formulation. Furthermore, it is seen that $\mathbf{B}^{b1}$ contains the second-order derivative of the shape function. Hence, it requires $C^1$-continuous approximation across inter-element boundaries in the finite element mesh. This is not a very trivial task in standard finite element method. In an effort to address this difficulty, several $C^0$ continuous elements [42-44] were then proposed. Alternatively, Hermite interpolation function with the $C^1$-continuity was added to satisfy specific approximation of transverse displacement [45]. They may produce extra unknown variables leading to an increase in the computational cost. It is now interesting to note that our present formulation based on NURBS basic functions satisfies naturally from the theoretical/mechanical viewpoint of FGM plates [46,47]. In our work, the basis functions are $C^{p-1}$ continuous. As a result, as $p \geq 2$, the present approach always satisfies $C^1$-requirement in approximate formulations based on the proposed higher order shear deformation theory.

## 4. Results and discussions

In this section, using XIGA, we study the natural frequency of the FGM plates with two kinds of crack: a center crack and an edge crack. Herein, ceramic-metal functionally graded plates of which material properties given in Table 1 are considered. In all numerical examples, we illustrate the present method using quadratic and cubic basis functions with full $(p+1) \times (q+1)$ Gauss points and the results, unless specified otherwise, are normalized as

$$\bar{\omega} = \omega \left( \frac{W^2}{h} \right) \sqrt{\frac{\rho_c}{E_c}} \tag{31}$$

Table 1: Material property.

|  | Al | ZrO$_2$ | Al$_2$O$_3$ |
|---|---|---|---|
| $E$ (GPa) | 70 | 200 | 380 |
| $\nu$ | 0.3 | 0.3 | 0.3 |
| $\rho$ (kg/m$^3$) | 2707 | 5700 | 3800 |

*4.1. Center crack plate*

First, convergence study of natural frequency of a thin isotropic cracked plate (ν=0.3) with length to thickness ratio $L/h$=1000 is considered. The plate with dimension $L \times W \times h$ has an initial crack at center as shown in Figure 6a. The comparison between the present method using TSDT and FSDT models and the analytical solution based on CPT by B. Stahl and LM. Keer [17] is depicted in Figure 7. It is observed that present method gives the upper convergence for natural frequency parameter. The TSDT model is more accurate than the FSDT: the first frequency changes from 2.83% to 0.63% for TSDT model and 5.35% to 1.08% in case of FSDT one according to increase in number of elements from 9x9 to 29x29.

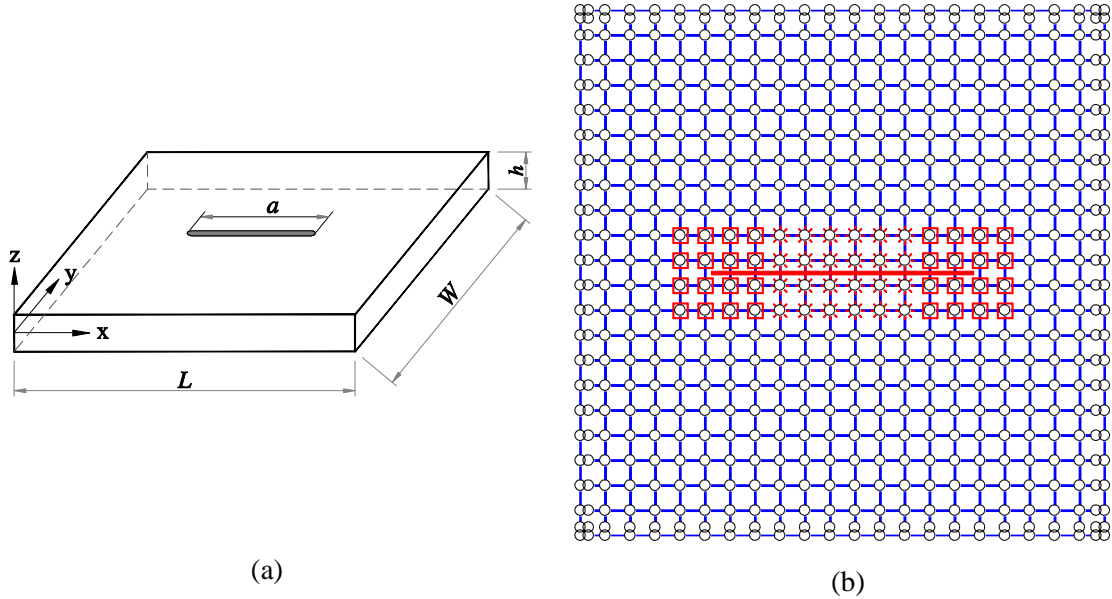

(a)　　　　　　　　　　　　　　　　　　　(b)

Figure 6. The plate with a center crack: (a) model; (b) mesh of 21x21 cubic NURBS elements.

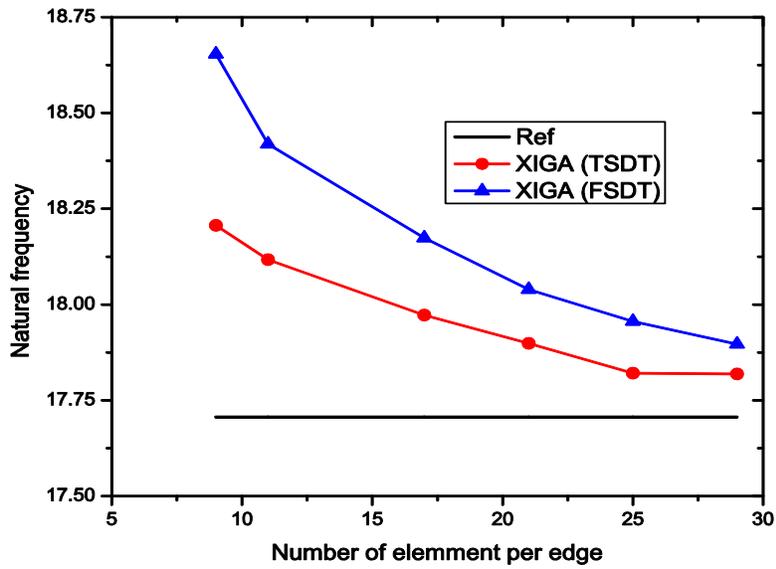

Figure 7. Convergence of the first frequency of the square plate with central crack with crack length ratio $a/L = 0.5$.

Next, the relation between non-dimension frequencies $\hat{\omega} = \omega L^2 \sqrt{\rho h / D}$ and crack length ratio according to mesh 21x21 as shown in Figure 6b is reported in Table 2. The obtained results from XIGA are in good agreement compared to both analytical solution [17, 18] and XFEM [48] using 20x20 nine-node Lagrange elements. For clearer vision, the comparison of first five frequencies between the present result and that of Stahl [17] and Liew et al. [18] is depicted in Figure 8. It is revealed that the frequencies decrease via increase in crack length ratio. For example, the values of frequency according to change of mode from 1 to 5 drop up to 18.3%, 67.3%, 5.3%, 40.8% and 23.7% of its initial values corresponding to an intact plate, respectively. It is concluded that the magnitude of the frequency according to anti-symmetric modes through the y-axis, which is perpendicular with cracked path (e.g. mode 2, mode 4, shown in Figure 9), is much more affected by the crack length. Here, the discontinuous displacement is shown clearly along crack path.

Table 2: Non dimensional natural frequency of an isotropic square plate with central crack

| Mode shape | Source | Crack length ratio $a/L$ | | | | | | |
|---|---|---|---|---|---|---|---|---|
| | | 0 | 0.2 | 0.4 | 0.5 | 0.6 | 0.8 | 1 |
| 1 | Stahl [17] | 19.7390 | 19.3050 | 18.2790 | 17.7060 | 17.1930 | 16.4030 | 16.1270 |
| | Liew [18] | 19.7400 | 19.3800 | 18.4400 | 17.8500 | 17.3300 | 16.4700 | 16.1300 |
| | XFEM [48] | 19.7390 | 19.3050 | 18.2780 | 17.7070 | 17.1800 | 16.4060 | 16.1330 |
| | XIGA(TSDT) | 19.7392 | 19.3846 | 18.4617 | 17.8989 | 17.3576 | 16.5157 | 16.1345 |
| | XIGA(FSDT) | 19.7398 | 19.4675 | 18.5933 | 18.0392 | 17.4949 | 16.6176 | 16.1355 |
| 2 | Stahl [17] | 49.3480 | 49.1700 | 46.6240 | 43.0310 | 37.9780 | 27.7730 | 16.1270 |
| | Liew [18] | 49.3500 | 49.1600 | 46.4400 | 42.8200 | 37.7500 | 27.4300 | 16.1300 |
| | XFEM [48] | 49.3480 | 49.1810 | 46.6350 | 43.0420 | 37.9870 | 27.7530 | 17.8260 |
| | XIGA (TSDT) | 49.3501 | 49.1906 | 47.1197 | 44.7124 | 39.3469 | 29.1186 | 16.1345 |
| | XIGA (FSDT) | 49.3831 | 49.2662 | 48.3935 | 46.4073 | 43.0769 | 33.2349 | 16.1355 |
| 3 | Stahl [17] | 49.3480 | 49.3280 | 49.0320 | 48.6970 | 48.2230 | 47.2560 | 46.7420 |
| | Liew [18] | 49.3500 | 49.3100 | 49.0400 | 48.7200 | 48.2600 | 47.2700 | 46.7400 |
|   | XFEM [48]  | 49.3480 | 49.3240 | 49.0320 | 48.6850 | 48.2140 | 47.2010 | 46.7340 |
|   | XIGA (TSDT)| 49.3501 | 49.3292 | 49.0903 | 48.6328 | 48.3547 | 47.3448 | 46.7376 |
|   | XIGA (FSDT)| 49.3831 | 49.3617 | 49.1684 | 48.9050 | 48.5105 | 47.5322 | 46.7790 |
| 4 | Stahl [17] | 78.9570 | 78.9570 | 78.6020 | 77.7330 | 75.5810 | 65.7320 | 46.7420 |
|   | Liew [18]  | 78.9600 | 78.8100 | 78.3900 | 77.4400 | 75.2300 | 65.1900 | 46.7400 |
|   | XFEM [48]  | 78.9550 | 78.9450 | 78.6000 | 77.7100 | 75.5790 | 65.7150 | 49.0990 |
|   | XIGA (TSDT)| 78.9589 | 78.9452 | 78.6507 | 77.6642 | 76.0779 | 67.2308 | 46.7376 |
|   | XIGA (FSDT)| 78.9999 | 78.9929 | 78.8136 | 78.4680 | 77.4980 | 71.7644 | 46.7790 |
| 5 | Stahl [17] | 98.6960 | 93.9590 | 85.5100 | 82.1550 | 79.5880 | 76.3710 | 75.2850 |
|   | Liew [18]  | 98.7000 | 94.6900 | 86.7100 | 83.0100 | 80.3200 | 76.6000 | 75.2800 |
|   | XFEM [48]  | 98.6980 | 93.8930 | 85.4500 | 82.1080 | 79.5560 | 76.3510 | 75.2750 |
|   | XIGA (TSDT)| 98.7292 | 94.5834 | 86.6987 | 82.7347 | 80.3835 | 76.7866 | 75.2823 |
|   | XIGA (FSDT)| 99.1816 | 95.9980 | 87.9910 | 84.2472 | 81.2667 | 77.3568 | 75.3936 |

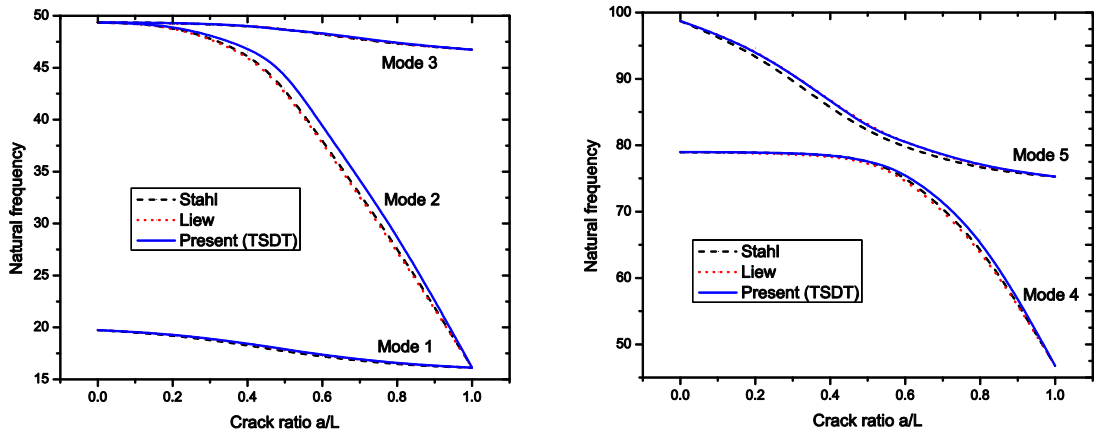

Figure 8. Variation of first five mode frequencies via length crack ratio.

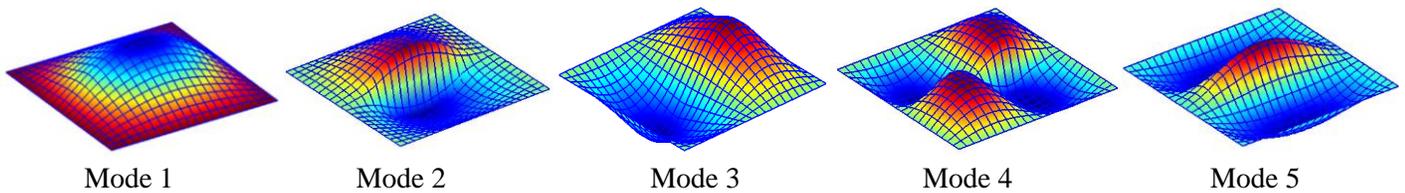

Mode 1      Mode 2      Mode 3      Mode 4      Mode 5

Figure 9. The first five mode shapes of simply supported plate having center crack with $a/L$ =0.8

Finally, the effect of length to thickness ratio $L/h$ on the frequencies of Al/Al$_2$O$_3$ FGM plate is revealed in Figure 10. In this example, the material properties are calculated by two homogenization schemes including the rule of mixture and the Mori-Tanaka scheme with the power index $n = 1$ according to Eqs. (2) and (4). With inhomogeneous material, the effective property through the thickness of the former is higher than the latter one. The results from the Mori-Tanaka scheme hence are lower than that of the counterpart because of lower stiffness. Moreover, the value of natural frequency changes rapidly between thick plate ($L/h$ =2) and moderate thin plate ($L/h$ =50) with discrepancy up to 57.2% and 71.3% via the rule of mixture and the Mori-Tanaka scheme, respectively. However, for thin plates ($L/h \geq 100$), it is independent on the length to thickness ratio with approximated difference up to 1% due to naturally shear-locking free by the present plate theory.

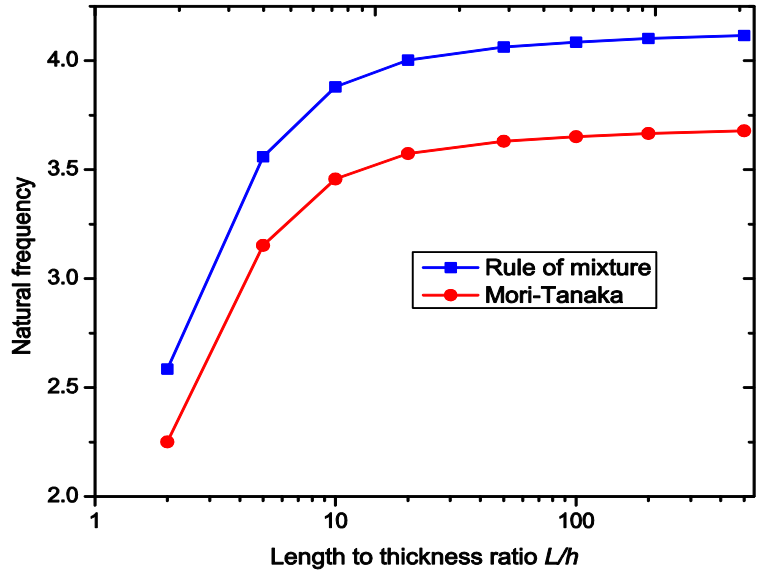

Figure 10. The first frequency of central cracked Al/Al$_2$O$_3$ plate computed with both of the rule of mixture and the Mori-Tanaka schemes.

*4.2. Edge crack plate*

Let us consider a square plate with uniform thickness $h$, length and width as $L$ and $W$, respectively. The plate, having a side crack from the left edge with length $a$, is discretized into 21x21 cubic NURBS elements as shown in Figure 11. Table 3 depicts the effect of the power index $n$ on the first five natural frequencies of a simply supported $Al/Al_2O_3$ plate with length to thickness ratio $L/h$=10 and crack length ratio $a/L$=0.5. The present method gives good agreement compared to both the Ritz method and XFEM. It can be seen that present results are slightly lower than that of Huang et al. [49] based on TSDT or even Natarajan et al. [26] based on FSDT. As compared to 3D elasticity solution calculated by ABAQUS finite element package (resulting in 130791 nodes) [49], the proposed approach based on TSDT model obtains the best first frequency. As shown in Figure 2, increasing volume fraction exponent $n$ reduces the effective property of the material through the plate thickness. Thus, frequency parameter $\bar{\omega}$ decreases respectively because of reduction the stiffness of FGM plate. The same conclusions are drawn for cantilever $Al/ZrO_2$ cracked plate with results listed in Table 4. Figure 12 and Figure 13 plot the first four mode shapes of the edge cracked FGM plate under full simply supported boundary and clamped edge right conditions, respectively. It can be seen that the magnitude of deflection based on anti-symmetric mode through the y-axis changes drastically around the crack path (for example, mode 2 and 4).

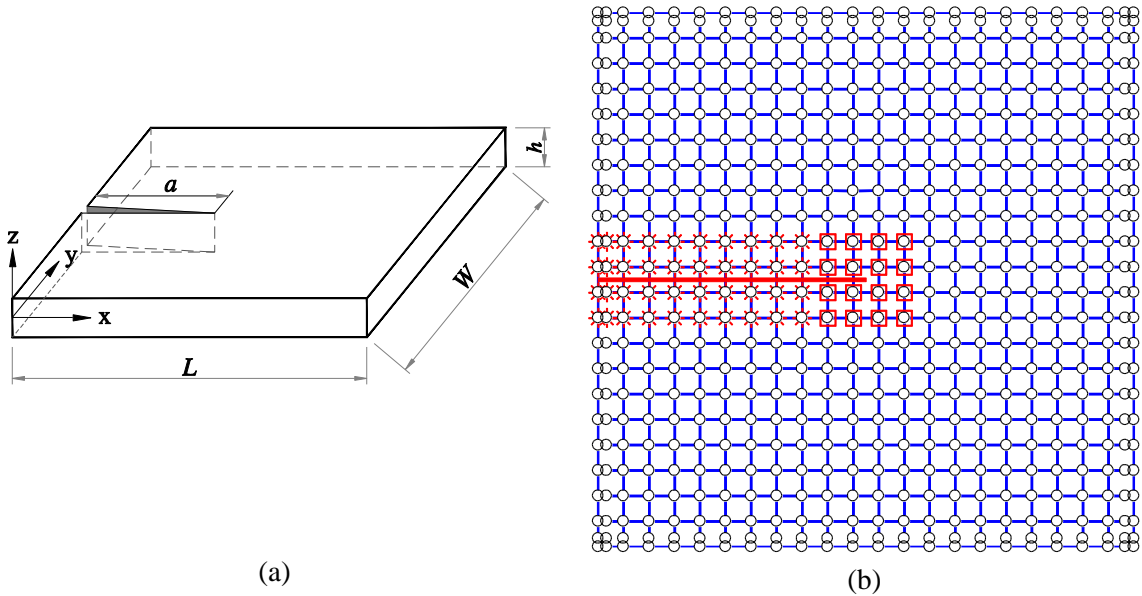

(a)        (b)

Figure 11. The plate with an edge crack: (a) model; (b) mesh of 21x21 cubic NURBS elements

Table 3: The first five frequencies of simply supported Al/Al$_2$O$_3$ plate with edge crack ($a/L$=0.5)

| $n$ | Method | Mode number | | | | |
|---|---|---|---|---|---|---|
| | | 1 | 2 | 3 | 4 | 5 |
| 0 | Ritz method [49] | 5.3790 | 11.4500 | 13.3200 | 16.1800 | 17.3200 |
| | XFEM [26] | 5.3870 | 11.4190 | 13.3590 | - | - |
| | XIGA (TSDT) | 5.3643 | 11.4734 | 13.2801 | 16.2062 | 17.2927 |
| | XIGA (FSDT) | 5.3657 | 11.3901 | 13.2818 | 16.2062 | 17.2433 |
| 0.2 | Ritz method [49] | 5.0010 | 10.6800 | 12.4100 | 15.4200 | 16.1500 |
| | XFEM [26] | 5.0280 | 10.6590 | 12.4370 | - | - |
| | XIGA (TSDT) | 4.9879 | 10.7069 | 12.3702 | 15.4377 | 16.1267 |
| | XIGA (FSDT) | 4.9877 | 10.6208 | 12.3641 | 15.4376 | 16.0678 |
| 1 | 3D elasticity [49] | 4.1150 | 8.8360 | 10.2400 | 13.3300 | 13.5200 |
| | Ritz method [49] | 4.1220 | 8.8560 | 10.2500 | 13.3100 | 13.4900 |
| | XFEM [26] | 4.1220 | 8.5260 | 10.2850 | - | - |
| | XIGA (TSDT) | 4.1119 | 8.8791 | 10.2131 | 13.3103 | 13.4946 |
| | XIGA (FSDT) | 4.1123 | 8.8129 | 10.2139 | 13.2728 | 13.4911 |
| 5 | Ritz method [49] | 3.5110 | 7.3790 | 8.6210 | 10.4900 | 11.1700 |
| | XFEM [26] | 3.6260 | 7.4150 | 8.5660 | - | - |
| | XIGA (TSDT) | 3.5018 | 7.3980 | 8.5912 | 10.4928 | 11.1511 |
| | XIGA (FSDT) | 3.5218 | 7.4559 | 8.6873 | 10.4956 | 11.2728 |
| 10 | Ritz method [49] | 3.3880 | 7.0620 | 8.2890 | 9.5690 | 10.7100 |
| | XFEM [26] | 3.4090 | 7.0590 | 8.2210 | - | - |
| | XIGA (TSDT) | 3.3773 | 7.0792 | 8.2582 | 9.5750 | 10.6887 |
| | XIGA (FSDT) | 3.3986 | 7.1420 | 8.3594 | 9.5757 | 10.8206 |

Table 4: The first five frequencies of cantilever Al/ZrO$_2$ plate with an edge crack ($a/L$=0.5)

| Mode | Method | $n$ | | | | |
|---|---|---|---|---|---|---|
| | | 0 | 0.2 | 1 | 5 | 10 |
| 1 | Ritz method [49] | 1.0380 | 1.0080 | 0.9549 | 0.9743 | 0.9722 |
| | XFEM [26] | 1.0380 | 1.0075 | 0.9546 | 0.9748 | 0.9722 |
|   | XIGA (TSDT)       | 1.0381 | 1.0076 | 0.9547 | 0.9738 | 0.9716 |
|   | XIGA (FSDT)       | 1.0380 | 1.0074 | 0.9546 | 0.9744 | 0.9721 |
| 2 | Ritz method [49]  | 1.7330 | 1.6840 | 1.5970 | 1.6210 | 1.6170 |
|   | XFEM [26]         | 1.7329 | 1.6834 | 1.5964 | 1.6242 | 1.6194 |
|   | XIGA (TSDT)       | 1.7363 | 1.6871 | 1.6006 | 1.6238 | 1.6189 |
|   | XIGA (FSDT)       | 1.7271 | 1.6778 | 1.5919 | 1.6191 | 1.6135 |
| 3 | Ritz method [49]  | 4.8100 | 4.6790 | 4.4410 | 4.4760 | 4.4620 |
|   | XFEM [26]         | 4.8231 | 4.6890 | 4.4410 | 4.4955 | 4.4845 |
|   | XIGA (TSDT)       | 4.8084 | 4.6782 | 4.4407 | 4.4743 | 4.4586 |
|   | XIGA (FSDT)       | 4.8015 | 4.6695 | 4.4340 | 4.4883 | 4.4693 |
| 4 | Ritz method [49]  | 5.2180 | 5.0780 | 4.8200 | 4.8500 | 4.8340 |
|   | XIGA (TSDT)       | 5.2332 | 5.0923 | 4.8336 | 4.8626 | 4.8457 |
|   | XIGA (FSDT)       | 5.2067 | 5.0644 | 4.8089 | 4.8618 | 4.8409 |
| 5 | Ritz method [49]  | 6.1850 | 6.0250 | 5.7160 | 5.5900 | 5.4780 |
|   | XIGA (TSDT)       | 6.1959 | 6.0246 | 5.7148 | 5.5986 | 5.4866 |
|   | XIGA (FSDT)       | 6.1950 | 6.0216 | 5.7139 | 5.5987 | 5.4866 |

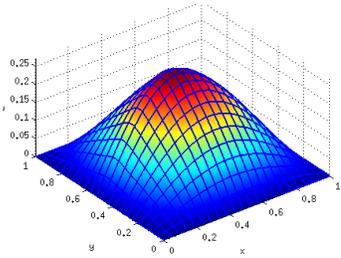 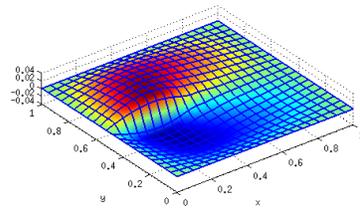 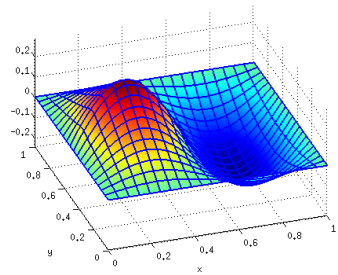 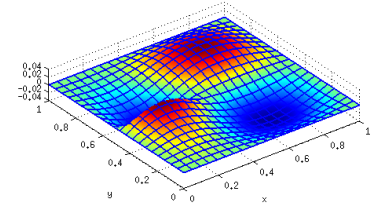

Mode 1　　　　　　　　Mode 2　　　　　　　　Mode 3　　　　　　　　Mode 4

Figure 12. First four mode shapes of simply supported Al/Al$_2$O$_3$ plate with edge crack.

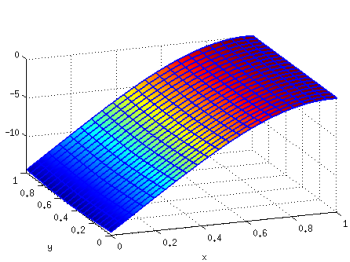 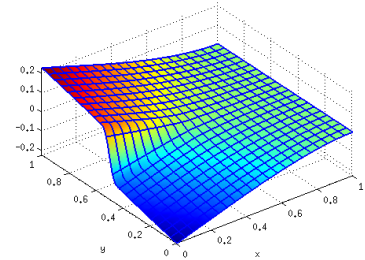 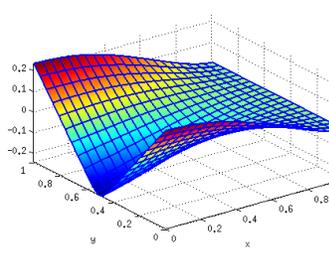 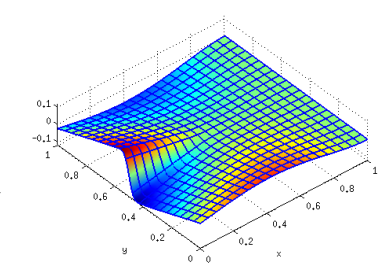

Mode 1　　　　　　　　Mode 2　　　　　　　　Mode 3　　　　　　　　Mode 4

Figure 13. First four mode shapes of cantilever Al/ZrO$_2$ plate with edge crack.

*4.3. Circular and annular plate with central crack*

In this example, plates with curved boundaries are studied, for example, annular geometry with uniform thickness *h*, outer radius *R* and inner one *r* as shown in Figure 14. The Al/Al$_2$O$_3$ FGM plate is clamped at the outer boundary and has centralized crack with length $a = (R-r)/2$. Herein, the Mori-Tanaka homogenization scheme is used.

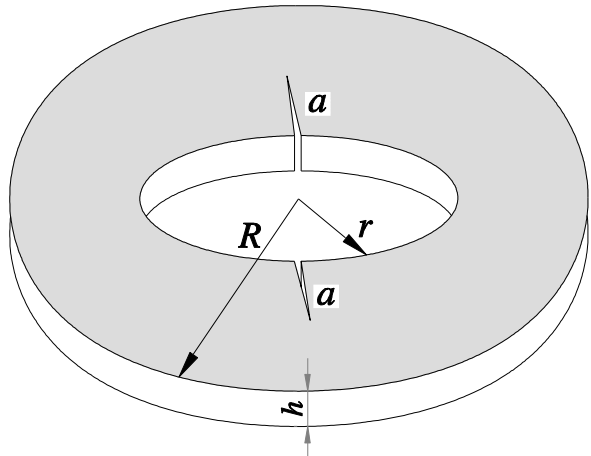

Figure 14. The model of annular plate.

By setting the inner radius $r = 0$, the model becomes the clamped circular plate with the central crack length 2*a*. Using *hk* refinement [30], we enable to mesh the domain into 21x21 quadratic and cubic NURBS elements as shown in Figure 15. It can be seen that the geometry of circular plate is described exactly based on NURBS basic functions.

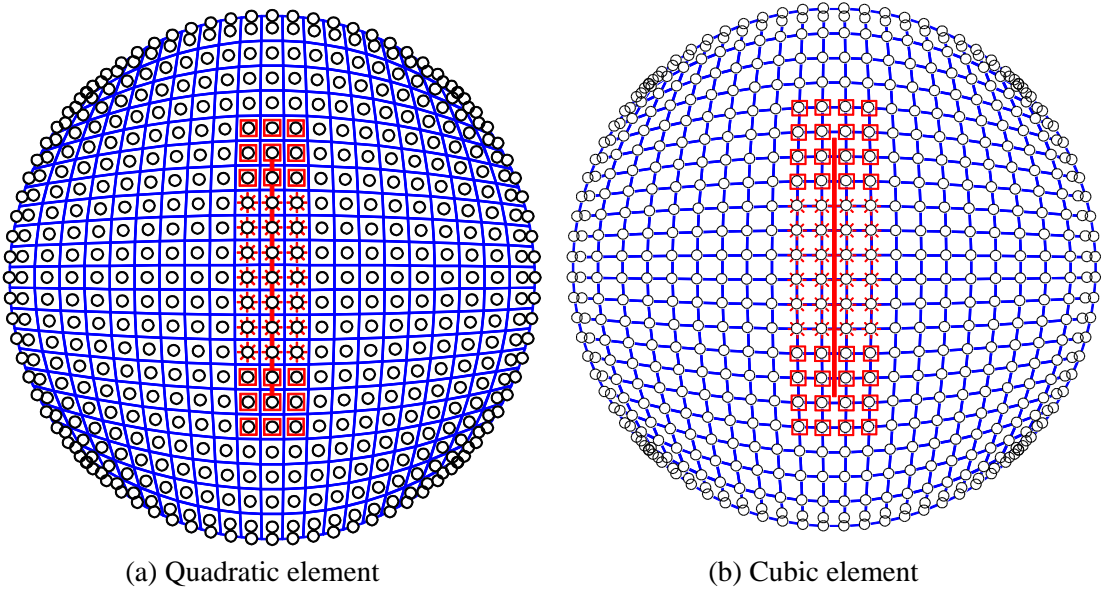

(a) Quadratic element  (b) Cubic element
Figure 15. Meshes of the circular plate with 21x21 NURBS elements.

Table 5 shows the first five natural frequency parameters via changing of the power index $n$ from 0 to 10. Observation is again that the value of frequency increases according to reduction of the value $n$. For this problem, XFEM is used to create reference results according to two meshes: 21x21 and 29x29 quadrilateral elements. It is shown that finer mesh is used lower results are obtained in this vibration problem. However, with the finer mesh (29x29), XFEM results are also higher than XIGA ones using quadratic element ($p=2$) or even cubic element ($p=3$). The differences may be caused by: (1) geometric error: curved geometry is exact description by XIGA based on NURBS instead of approximation in XFEM; (2) approximated order: XIGA utilizes NURBS with higher order functions than XFEM using bilinear Lagrange functions; (3) plate model: a similar tendency was also concluded in Ref [47] in analyzing vibration of $Al/Al_2O_3$ plate. HSDT used in proposed model causes more flexible stiffness matrix than FSDT which is suitable to $C^0$ continuous element in XFEM. Figure 16 plots the mode shapes of the circular plate. Using NURBS functions, the curved boundary of the circular plate is still described exactly.

Table 5: The first five frequencies of clamped circular Al/Al$_2$O$_3$ plate based on meshing of 21x21 elements with the central crack ($a/R = 0.5$)

| $n$ | Method | Mode number | | | | |
|---|---|---|---|---|---|---|
| | | 1 | 2 | 3 | 4 | 5 |
| 0 | XFEM | 2.8864 | 5.0626 | 6.4816 | 9.4784 | 9.8883 |
| | | (2.7650) | (4.7604) | (6.2044) | (9.0369) | (9.5031) |
| | XIGA ($p=2$) | 2.6492 | 4.7090 | 5.9402 | 8.7118 | 9.2582 |
| | XIGA ($p=3$) | 2.6288 | 4.3852 | 5.8220 | 8.5107 | 8.8251 |
| 0.2 | XFEM | 2.4278 | 4.2749 | 5.4563 | 7.9876 | 8.3275 |
| | | (2.3184) | (4.0077) | (5.2068) | (7.5894) | (7.9838) |
| | XIGA ($p=2$) | 2.2132 | 3.9524 | 4.9667 | 7.2896 | 7.7550 |
| | XIGA ($p=3$) | 2.1958 | 3.6785 | 4.8698 | 7.1194 | 7.4010 |
| 1 | XFEM | 1.9914 | 3.5128 | 4.4755 | 6.5504 | 6.8313 |
| | | (1.9005) | (3.2907) | (4.2685) | (6.2203) | (6.5464) |
| | XIGA ($p=2$) | 1.8117 | 3.2485 | 4.0639 | 5.9662 | 6.3444 |
| | XIGA ($p=3$) | 1.7963 | 3.0089 | 3.9821 | 5.8150 | 6.0484 |
| 5 | XFEM | 1.7717 | 3.0949 | 3.9651 | 5.7828 | 6.0356 |
| | | (1.7037) | (2.920) | (3.8097) | (5.5346) | (5.8165) |
| | XIGA ($p=2$) | 1.6347 | 2.8932 | 3.6457 | 5.3306 | 5.6442 |
| | XIGA ($p=3$) | 1.6212 | 2.6789 | 3.5682 | 5.1932 | 5.3636 |
| 10 | XFEM | 1.6935 | 2.9525 | 3.7880 | 5.5230 | 5.7635 |
| | | (1.6303) | (2.7887) | (3.6433) | (5.2917) | (5.5586) |
| | XIGA ($p=2$) | 1.5677 | 2.7692 | 3.4969 | 5.1124 | 5.4135 |
| | XIGA ($p=3$) | 1.5552 | 2.5694 | 3.4236 | 4.9850 | 5.1474 |

The results in parenthesis according to meshing of 29x29 quadrilateral elements.

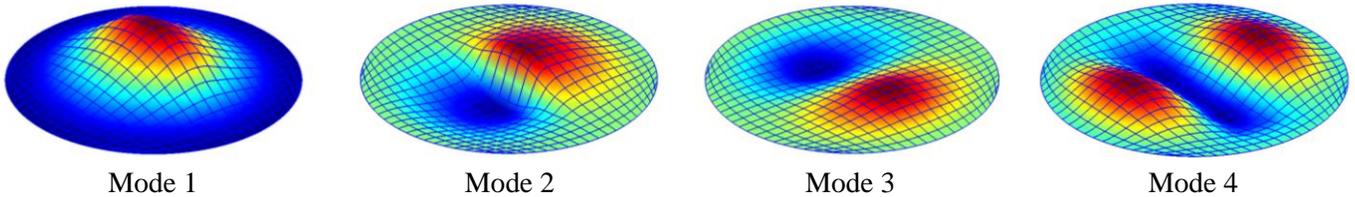

Mode 1  Mode 2  Mode 3  Mode 4

Figure 16. First four mode shapes of the clamped circular plate with central crack

As the inner radius $r \neq 0$ we have the full annular plate shown in Figure 14. Because of symmetry, an upper haft of plate has been modeled in Figure 17 with the symmetric constraint: displacement along *y*-direction equals to zero at *y* = 0. Based on Eq. (5) the conditions for symmetric boundary are:

$$v_0 = \beta_y = w_{,y} = 0 \tag{32}$$

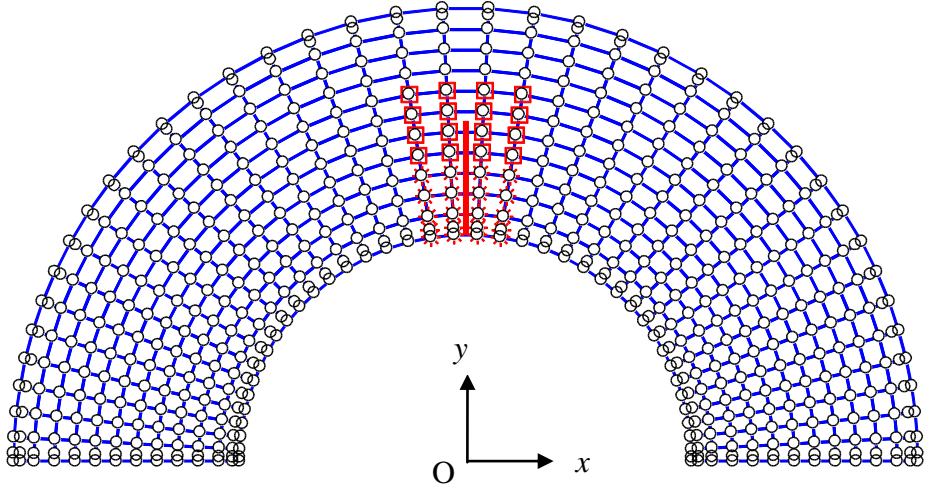

Figure 17. Mesh of an upper half of annular plate.

The Dirichlet boundary conditions $v_0 = \beta_y = 0$ can be enforced easily. However, the condition $w_{,y} = 0$ is solved in a different way follow a stream function formulation [50]. The derivative can be included in a compact form of the normal slope at the boundary:

$$\frac{\partial w}{\partial y} = \lim_{\Delta y \to 0} \frac{w(y_C + \Delta y) - w(y_C)}{\Delta y} \tag{33}$$

Eq. (32) leads to impose the same deflection of two rows of control points: which lies on symmetric edges and its adjacent row.

Table 6 tabulates the frequency parameter of the annular Al/Al$_2$O$_3$ plates via outer radius to inner radius ratio *R/r* and radius to thickness ratio *R/h* according to *n*=1. It is concluded that the frequency parameters decrease sequentially following to increase in inner radius to outer radius ratio *r/R*. To enclose this section the first four mode shapes of annular FGM plate are depicted in Figure 18.

Table 6: The frequency parameter $\tilde{\omega} = \omega(R-r)^2 / h\sqrt{\rho_c / E_c}$ of annular pate via inner radius to outer radius ratio *r/R* and radius to thickness ratio *R/h* according to *n*=1.

| R/h | r/R | Mode number | | | | |
|---|---|---|---|---|---|---|
| | | 1 | 2 | 3 | 4 | 5 |
| 2 | 0 | 1.2786 | 1.7682 | 2.3336 | 2.7352 | 2.8058 |
| | 0.2 | 0.8438 | 1.0109 | 1.7316 | 1.8588 | 1.9021 |
| | 0.5 | 0.5516 | 0.5896 | 0.7308 | 0.8458 | 0.9817 |
| | 0.8 | 0.2760 | 0.2771 | 0.2800 | 0.2896 | 0.2905 |
| 5 | 0 | 1.6804 | 2.6230 | 3.5290 | 4.9598 | 5.0683 |
| | 0.2 | 1.0877 | 1.3898 | 2.7728 | 3.4371 | 4.0295 |
| | 0.5 | 0.7845 | 0.8556 | 1.1536 | 1.3664 | 1.9105 |
| | 0.8 | 0.4923 | 0.4965 | 0.5066 | 0.5225 | 0.5537 |
| 10 | 0 | 1.8480 | 3.5185 | 4.0473 | 5.9916 | 6.3512 |
| | 0.2 | 1.1563 | 1.5352 | 3.1932 | 4.1404 | 4.8849 |
| | 0.5 | 0.8621 | 0.9560 | 1.3533 | 1.6388 | 2.3101 |
| | 0.8 | 0.6470 | 0.6540 | 0.6730 | 0.6975 | 0.7545 |
| 20 | 0 | 1.8379 | 3.1941 | 4.1533 | 6.1526 | 6.4956 |
| | 0.2 | 1.1793 | 1.598 | 3.3309 | 4.4139 | 5.2045 |
| | 0.5 | 0.8877 | 0.9954 | 1.4507 | 1.7937 | 2.5902 |
| | 0.8 | 0.7279 | 0.7371 | 0.7655 | 0.7999 | 0.8775 |
| 100 | 0 | 1.8649 | 3.3264 | 4.2398 | 6.3270 | 6.7435 |
| | 0.2 | 1.1922 | 1.6442 | 3.4202 | 4.6187 | 5.3415 |
| | 0.5 | 0.8973 | 1.0154 | 1.5007 | 1.8656 | 2.6187 |
| | 0.8 | 0.7646 | 0.7753 | 0.8124 | 0.8567 | 0.9362 |

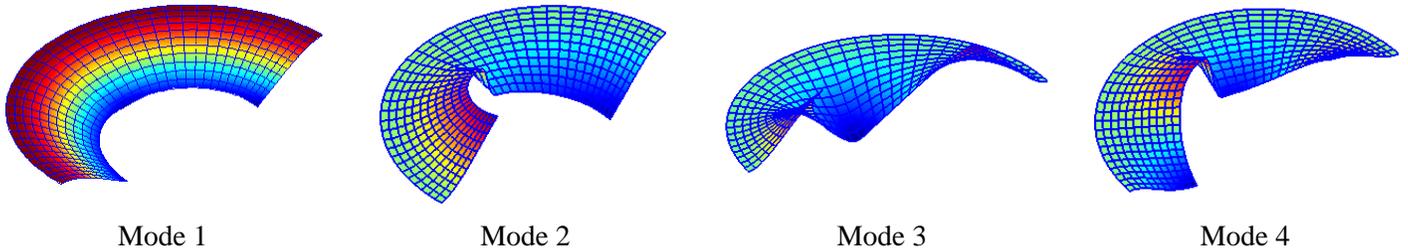

Mode 1        Mode 2        Mode 3        Mode 4

Figure 18. The first four mode shapes of the annular plate with *R/r*=2, *R/h*=10.

## 5. Conclusions

In this paper, a novel and effective formulation based on combining XIGA and HSDT has been applied to dynamic analysis of the cracked FGM plates. The present method utilizing NURBS basis functions allows us to achieve easily the smoothness with arbitrary continuous order compared with the traditional FEM. Consequently, it naturally fulfills the $C^1$-continuity of HSDT model which is not dependent on dispersed SCF. Furthermore, the special enrichment functions are applied to describe the singularity behaviors of the cracked plates. The obtained results in excellent agreement with that from analytical and numerical methods in the literature demonstrate that XIGA is an effectively computational tool for vibration analysis of the cracked plates.

It is also concluded that magnitudes of the natural frequency decrease via increase in crack length ratio. They change dramatically according to anti-symmetric mode through the *y*-axis which is perpendicular with crack path.

Herein, two homogenous models based on exponent function of *n* have been used to estimate the effective property of the FGM plates include the rule of mixture and the Mori-Tanaka technique. It can be seen that, increasing power index *n* leads to a reduction of frequency parameter of the FGM plates. In addition, to consider the interactions among the constituents, the Mori-Tanaka homogenization scheme gains lower frequency value than the rule of mixture.

Besides, study the benchmarks in rectangular geometry for purpose of comparison, extensive studies are carried out for circular and annular FGM plates. It is believed that XIGA with non-geometric approximation can be very promising to provide the good reference results for vibration analysis of these plates with curved boundaries.

**Acknowledgement**

VPN would like to acknowledge the partial financial support of the Framework Programme 7 Initial Training Network Funding under grant number 289361 "Integrating Numerical Simulation and Geometric Design Technology".